\documentstyle[aps,preprint]{revtex}

\tighten

\newcommand {\R} {\mbox{I\hspace{-.4ex}R}}
\newcommand{\id}{\mbox{1\hspace{-.2ex}\rule{.1ex}{1.44ex}}
    \hspace{-.82ex}\rule[-.01ex]{1.07ex}{.1ex}
    \hspace{-.7ex}\rule[1.44ex]{.4ex}{.06ex}\hspace{.2ex}}
\newcommand {\Tr}   {\,\mbox{\rm Tr}\,}
\newcommand {\NLSM}  {NL$\sigma$M}
\newcommand {\NLSMs} {NL$\sigma$Ms}
\newcommand {\RP}[1]{\R P^{#1}}

\begin{document}
%
\title{Importance of the topological defects for two dimensional phase
transitions
       and their relevance for the renormalization group}
\author{ Gil Zumbach}
\address{ Harvard University, Physics Department, Cambridge MA 02138, USA \\
          e-mail: Zumbach@cmt.harvard.edu}
\date{\today}
\maketitle
%
\begin{abstract}
For various two dimensional non linear $\sigma$ models,
we present a direct comparison between the $\beta$ functions computed with the
$2+\epsilon$ renormalization group
and the $\beta$ functions measured by Monte Carlo simulations.
The theoretical and measured $\beta$ functions match each other nicely for
models with a trivial topology,
yet they disagree clearly for models containing topological defects.
In these later cases, they are compatible with a phase transition at a finite
temperature.
This indicates that the global properties of the manifold do matter,
in contradiction with the assumption used in the $2+\epsilon$ RG computation.
\end{abstract}
%

In this letter, we want to address problems related to phase transitions in two
dimensions ($d=2$)
(for models with continuous symmetries),
the various non linear $\sigma$ models (\NLSM ), the renormalization group
(RG),
and the possible role of topological defects.
First, we will review carefully our theoretical understanding of the various
points just mentioned.
Then, we will present extensive Monte Carlo simulations for different \NLSM s,
and in particular a direct comparison with the RG $\beta$ function at $d=2$.

The most studied models of phase transitions with continuous symmetry
are the $S^{n-1}$ Heisenberg models (or the $n$-components $\varphi^4$ models,
as they are widely believed to be in the same universality class).
In two dimensions, the accepted picture is that they have a phase transition at
$T_c=0$,
with a correlation length diverging as $\xi \sim \exp(c/T)$
(we will discuss below the special case of the $S^1$ or X-Y model).
This picture has been extended to essentially every two dimensional models with
continuous symmetry.
Of particular interest are the non linear $\sigma$ models.
In a  \NLSM, the field has values in a manifold $M=G/H$,
taken as the coset of a Lie group $G$ by a subgroup $H$.
These models are constructed so as to have a phase transition from a high
temperature phase
with symmetry $G$ broken down to a low temperature phase with symmetry $H$.
Their importance comes from the fact that, if there is universality in this
larger class of models,
the \NLSM\ is the canonical model to describe a phase transition with symmetry
$G$ broken down to $H$.
The \NLSMs\ correspond to the limit where all massive modes are neglected.
The simplest example of a \NLSM\ is constructed with $G=O(n)$ broken down to
$H=O(n-1)$,
the manifold $M = O(n)/O(n-1) = S^{n-1}$ is the $n-1$ dimensional sphere.
A point on a unit sphere is conveniently described by a $n$-components unit
vector,
and we find again the usual $S^{n-1}$ Heisenberg model.

Polyakov \cite{Polyakov} initiated the RG computations in $d=2+\epsilon$ for
the Heisenberg models.
The method was then further applied to all \NLSMs \cite{Friedan}.
The basic idea is to expand the Hamiltonian in a spin waves perturbation in the
low temperature phase,
i.e. to expand in the small displacements around an ordered state.
Clearly, this method probes only the local structure of the manifolds.
This is embodied by the resulting $\beta$ function, which depends only
on the curvature of the manifolds (at two loops order), a purely local
geometrical quantity.
In two dimensions, the computations predict generically a phase transition at
$T_c=0$,
with a correlation length diverging exponentially fast.
However, the basic idea may appear dubious,
because the transition between
a low temperature phase (which depends only on the local properties of $M$)
and a high temperature phase (which depends on the global properties of $M$)
ought not to depend only on the local properties of the manifold.
Strictly in two dimensions, the low temperature phase is even nonexistent, and
yet,
using this fictitious phase, we obtain predictions in the high temperature
phase!
The importance of the $2+\epsilon$ RG computations comes from the fact that,
except for the Heisenberg case,
we do not have that many other reliable methods to obtain information about
phase transitions
(see e.g. \cite{Gil_94_II} for a discussion of the problems related to the
computations
from the upper critical dimension and for the large $n$-expansion for the
Stiefel model).
To make it worse, we are often interested in three dimensional systems,
or in quantum systems, i.e. to take $\epsilon =1$ in the $d=2+\epsilon$
expansion.
Therefore, we need to have a reliable understanding of what happens at
$\epsilon =0$.
Finally, let us note that, if the perturbative RG gives a prediction
for the asymptotic regime (i.e. $\xi\rightarrow\infty$),
it gives no estimate for the correlation length (or temperature) at which the
asymptotic regime is reached.

The role of topological defects in two dimensions is a related lingering
problem.
Most of our comprehension comes from the X-Y or $S^1$ Heisenberg model.
This model is unique for several reasons.
The main one is that the manifold is of dimension one,
and therefore possesses no curvature.
This absence of curvature arises because the group $G\sim O(2)$ is Abelian.
The related spin waves $\beta$ function is then identically zero.
Yet, this model possesses defects as $\Pi_1(S^1) = Z_2$,
and this drives the Berezinskii-Kosterlitz-Thouless transition \cite{BKT}.
Another peculiarity of the X-Y model is the Villain approximation
\cite{Villain},
which decouples spin waves and defects,
allowing to build a picture of the phase transition in terms of the defects
only.
Other \NLSM s have topological defects,
as the related manifolds have some non trivial homotopy groups.
Yet all models other than the X-Y model have curvatures.
It has not been possible, up to now,
to build an effective Hamiltonian for the spin waves and the topological
defects
(with their interactions) starting from the bare models.
However, according to the RG argument, the spin waves alone are enough to
disorder the system,
and the topological defects are simply irrelevant.
Thus, the \NLSMs\ offer a larger testing ground for two dimensional phase
transitions, the RG,
and for the role of topological defects,
away from all the peculiarities of the $S^1$ Heisenberg model.

Essentially, the only tool at our disposal to test the various theoretical
ideas
presented above is the Monte Carlo (MC) simulation.
In this letter, we present a direct comparison between the $\beta$ function
computed with the RG
and the results of MC simulations.
The basic idea is quite simple.
Let us consider a model with cut-off $\Lambda$ and one coupling constant,
called $T$.
The correlation length $\xi(T, \Lambda)$ obeys the RG equation
$(\Lambda\partial_\Lambda + \beta_T \partial_T)\xi = 0$.
As the correlation length is of dimension -1 we have
$\Lambda\partial_\Lambda\xi = -\xi$ and
the previous relation can be put into the form $\beta_T = 1/\partial_T \ln
\xi$.
Then, the correlation length can be measured in a MC simulation,
from which we can compute $1/\partial_T \ln \xi$ and compare with the RG
$\beta_T$ function.
Let us emphasize that there are {\em no adjustable parameters } in this
process.
The difficulty lays in the MC simulations because the correlation length is a
poorly self averaging quantity,
from which we have to take the logarithm and then a derivative.
This will amplify the noise, and therefore we will need to have outstanding
statistics in the MC simulations.

Having presented the basic idea, let us turn to the general case.
For a model defined with a cut-off $\Lambda$ and a set of coupling constants $g
= \{ g_i \}$,
the (bare) $n$ points vertex functions satisfy a Callan-Zymanzik equation
 \begin{equation}
  \left[ \Lambda\partial_\Lambda + \beta_i \partial_{g_i} - \frac{n}{2}\eta
\right]
    \Gamma^{(n)}(k_j; g, \Lambda) \simeq 0
\end{equation}
with $\beta_i(g) = \Lambda\partial_\Lambda g_i$, $\eta(g) =
\Lambda\partial_\Lambda \ln Z$,
and $Z$ is used to renormalize the field.
The right hand side is not strictly zero but it depends on a power of the
cut-off,
and is neglected according to a text book argument.
This relation can be put into the form
$\frac{n}{2}\eta = \left[ \Lambda\partial_\Lambda + \beta_i
\partial_{g_i}\right]\ln \Gamma^{(n)}$.
{}From the bare two points vertex function,
we can compute the susceptibility $\chi^{-1} = \Gamma^{(2)}(k=0)$
and the correlation length $\xi^2 = \left.\partial_{k^2} \ln
\Gamma^{(2)}\right|_{k=0}$.
Combining these equations, we obtain a (Callan-Zymanzik) equation for the
correlation length
$[\Lambda\partial_\Lambda + \beta_i \partial_{g_i} ]\xi= 0$ and for the
susceptibility
$[\Lambda\partial_\Lambda + \beta_i \partial_{g_i} + \eta]\chi = 0$.
The operator $\Lambda\partial_\Lambda$ measures the dimension of the object it
is applied on,
and $\Lambda\partial_\Lambda \ln\xi = -1$, $\Lambda\partial_\Lambda \ln\chi =
-2$.
Inserting in the above equation, we obtain our starting relation
 \begin{eqnarray}
  \beta_i \partial_{g_i} \ln \xi  & = & 1  \\
  \beta_i \partial_{g_i} \ln \chi & = & 2 - \eta(g)  \nonumber
 \end{eqnarray}
For a model with one coupling constant $T$, this becomes
 \begin{eqnarray}
  \beta_T   & = & 1/ \partial_T \ln \xi  \\
  \eta(T) & = & 2 - \frac{\partial \ln \chi}{\partial \ln \xi}
 \end{eqnarray}
For a model with two coupling constants $T$ and $\alpha$, we obtain for the
correlation length
 \begin{equation}
  \frac{\beta_T}{1-\beta_\alpha \partial_\alpha\ln \xi} = \frac{1}{\partial_T
\ln \xi}
 \end{equation}
In these relations, the left hand side can be computed with the RG,
and the right hand side can be measured by MC simulations.

Let us now define the \NLSM s we are working with.
The space is a square lattice of size $L^2$, with periodic boundary conditions.
On each site $x$ of the lattice,
there is a field $\varphi(x)$ with values in a manifold.
The first set of models is the Stiefel model $V_{n,p}$,
corresponding to the group $G=O(n)$, the subgroup $H=O(n-p)$
and the manifold $V_{n,p} = O(n)/O(n-p)$ \cite{Gil_94_II}.
The field $\varphi$ is a $n\times p$ matrix, with the constraint
$\varphi^+\varphi = \id_p$.
The field can also be viewed as $p$ orthonormal $n$-components vectors
$\varphi_i$,
or as a $p$ frame in $\R^n$.
The Hamiltonian is
 \begin{equation}
    H =  -\sum_{<x,y>} \Tr \varphi^+(x)\; \varphi(y)
      =  -\sum_{<x,y>} \sum_i \varphi_i(x)\cdot \varphi_i(y)
 \end{equation}
and the sum is over each link of the lattice.
For $p=1$, this is just the $n$-components Heisenberg model $S^{n-1} \sim
V_{n,1}$.
Let us note the special cases $V_{n, n-1} \sim SO(n)$ and $V_{n,n} \sim O(n)$.
We will also use the $SO(3)$ chiral model, corresponding to $n=p=3$,
but where the field obeys the supplementary constraint $\det \varphi = +1$.
The second set of models is the Grassmann $\RP{n-1}$ model,
corresponding to the group $G=O(n)$ and the subgroup $H = O(1)\times O(n-1)$.
The field is a $n$-components unit vector, and the Hamiltonian is
 \begin{equation}
    H = -\sum_{<x,y>} \left(\varphi(x)\cdot \varphi(y)\right)^2
 \end{equation}
We made extensive MC simulations for all models with a manifold
of dimension $D=1$, namely $V_{2,1}\sim S^1$ (or X-Y model), $V_{2,2} \sim
O(2)$, and
of dimension $D=3$, namely $S^3$, $\RP{3}$, $V_{3,2} \sim SO(3)$, $V_{3,3} \sim
O(3)$.
We also made simulations for the $D=7$ model $V_{5,2}$.

There exists an isomorphism between the manifolds $\RP{3} \sim V_{3,2} \sim
SO(3)$.
The Hamiltonian are however different,
except for the $\RP{3}$ model which can be mapped into the $SO(3)$ chiral model
\cite{KZ93}.
There are also the {\em local } isomorphisms between the manifolds $S^1 \sim
O(2)$
and $S^3 \sim \RP{3}\sim V_{3,2} \sim SO(3) \sim O(3)$.
Concerning the global properties,
the homotopy groups are $\Pi_0(O(2)) = \Pi_0(O(3)) = Z_2$ (boundary defect or
domain walls),
$\Pi_1(\RP{3}) = \Pi_1(V_{3,2}) = Z_2$ and $\Pi_1(S^1) = Z$ (point defects).
For the manifolds $S^3$ and $V_{5,2}$,
the first homotopy groups $\Pi_0$, $\Pi_1$ and $\Pi_2$ are trivial
(no topological defects of any kind, including instanton).

Let us now turn to the RG $\beta$ function for these models.
For the $S^{n-1}$ models, we have the 5 loops results of Wegner \cite{Wegner}.
As the manifolds $\RP{n-1}$ are locally identical to $S^{n-1}$,
the models have an identical $\beta$ function (at all order),
with the following relation between the temperatures $T(\RP{n-1}) = 2
T(S^{n-1})$.
For the $O(3)$ model, as the low temperature continuum limit neglects the two
sheets of $O(n)$,
it is identical to the $SO(3)$ chiral model, and hence,
via the previously mentioned mapping, to the $\RP{3}$ model.
Therefore, the $\beta$ function is the $S^3$ $\beta$ function (again, at all
order),
with the relation $T(O(3)) = 8 T(S^3)$.
For the $V_{n,2}$ models, the $\beta$ functions have been computed by
various authors \cite{KZ93,ADJ,Kawamura_91} and \cite{ADDJ} where the
computations are more detailed.
In the continuum limit, the action can be parametrized as
\begin{equation}
  \beta H = \frac{1}{2T}\frac{2}{1+\alpha} \int d^2x\;         \left[
                    (\nabla\varphi_1)^2 + (\nabla\varphi_2)^2             %
                  + (\alpha-1)(\varphi_1\nabla\varphi_2)^2
                     \rule{0mm}{3ex}\right]                                %
 \end{equation}
with the two coupling constants $T$ and $\alpha$.
The bare models correspond to $\alpha = 1$.
At one loop order, the $\beta$ functions in $d=2+\epsilon$ are
 \begin{eqnarray}
  \beta_T    & = & \epsilon T - (n-2) \left( \frac{1+\alpha}{2} \right)^2
\frac{T^2}{2\pi}
        \label{beta_T_alpha} \\
  \beta_\alpha & = & \frac{T}{2\pi}   \left( \frac{1+\alpha}{2} \right)^2
\left\{ (n-1)\alpha - (n-3)\right\}.
 \end{eqnarray}
For $d=2$, a trajectory of the flow possesses the invariant
  $K = T |(n-1)\alpha -(n-3) |^{(n-2)/(n-1)}$.
Then, we can integrate by the method of characteristics the flow equation and
the differential equation for $\xi$.
With the value $\alpha = 1$ corresponding to the bare $V_{n,2}$ models, we
obtain
\begin{equation}
\left.\frac{\beta_T}{1-\beta_\alpha\partial_\alpha \ln \xi}\right|_{\alpha = 1}
=
   - \frac{T^2}{2\pi}\;\frac{n-2}
    {1+\frac{n-1}{n-2}\sum_{k=0}\frac{1}{(n-1)k + n -2
}\left(\frac{-1}{n-2}\right)^k}
  \label{beta_V_n2}
\end{equation}
By comparing eq.~(\ref{beta_V_n2}) with eq.(\ref{beta_T_alpha}),
we see that the term $1-\beta_\alpha\partial_\alpha \ln\xi$ changes only the
proportionality constant in $\beta_T$.
For $n=3$, we have computed the above function at 2 loops order.
In the domain of temperature we are interested in,
the two loops correction is of order $2 \sim 3$\% and therefore can be
neglected.
For the $S^3$ model, in the range of correlation length accessible to MC
simulations,
the 5 loops result gives a correction of order 10\% over the one loop result.
Similarly, for the $\RP{3}$ and $O(3)$ models,
the 5 loops $\beta$ function gives a correction of order 2\%.
Clearly, the lowest order $\beta$ function is already a good estimate for all
models.

The Monte Carlo program has already been described in \cite{KZ93}.
Briefly, we use the Ferrenberg-Swendsen multi-histograms method \cite{FS}
and a cluster update algorithm \cite{Wolff}, eventually biased according to
\cite{Gil_biais}.
The correlation length is computed based on the small $k$ behavior of the
Fourier transformed
two points Green function $\tilde{G}(k)$.
{}From the histograms, we get an estimate for $\xi(T)$ from which we compute
$1/\partial_T\ln\xi$.
Let us emphasize again that we need very good statistics in order
to obtain a satisfactory estimate for this last quantity.
The total computation time for the present simulations is within the range of
several years on a typical RISC workstation.
The largest size for which we can attain good enough statistics is typically
$L=128$.
For the renormalization of the field $\eta(T)$, we can obtain a rough estimate,
but our statistics are still insufficient to iron out the wrinkles on the
curves.
Finally, even with a cluster update, the autocorrelation time grows badly
around the transition.

Due to a lack of space, we can present the graphical results for only three
representative models
(by ``representative'' we mean that the $\beta$ functions look similar).
Let us emphasize that there are no adjustable parameters in these results,
neither in the MC measurements nor in the theoretical RG $\beta$ functions.
Figure 1 presents the results for the $V_{5,2}$ model, and is also
representative of the $S^3$ model.
The low temperature asymptotic regime is entered for correlation length of
order $\xi \sim 10$,
and the measured and computed $\beta$ functions match nicely.
To the best of our knowledge, this is the first check of the $2+\epsilon$ RG
idea
for a model different from the Heisenberg model.
In the $V_{5,2}$ and $S^3$ models, there are no topological defects.

Figure 2 presents the results for the $\RP{3}$ model
and is also representative of the $V_{3,2}$ and $S^1$ model.
There we can see the clear discrepancy between the RG prediction and the MC
measurement.
The measured $\beta$ function goes even below the RG prediction,
a disturbing fact (although it does not violate any known rigorous bound).
More disturbing is the fact that for $T(\RP{3}) \simeq 0.3$,
the equivalent $S^3$ temperature is $T(S^3) = T(\RP{3})/2 \simeq 0.15$,
a temperature way inside the asymptotic regime with a correlation length of
order $\xi(S^3, T=0.15) \simeq 10^8$.
This series of models $\RP{3}$, $V_{3,2}$ and $S^1$ contains point defects as
$\Pi_1 \neq 0$.
In particular, for the $V_{3,2}$ model used to describe frustrated
antiferromagnets,
this should be a clear warning (not to speak about the $\epsilon=1$
extrapolation \cite{ADJ}).

Figure 3 presents the results for the $O(3)$ model, and is also representative
of the $O(2)$ model.
The discrepancy between the RG and MC $\beta$ function appears even more
obviously.
Again, through the equivalence with $S^3$ in the continuous limit,
the temperature $T(O(3)\simeq 1$ ends up way inside the asymptotic region for
the $S^3$ model,
also with a correlation length of order $\xi \simeq 10^8$.
The $O(2)$ and $O(3)$ models have boundary defects as $\Pi_0(O(n)) = Z_2$.

{}From our grouping, the key factor appears to be the various topological
defects.
In our MC simulations, we also measured the density of the different defects
and the transition is always clearly related to them.
Obviously, there is a problem with the $d=2$ RG $\beta$ function for models
with topological defects.
In contradiction with the RG spin waves assumption,
{\em the global properties of the manifold do matter}
and a topological term seems to be a relevant direction for the RG flow.
In fact, in view of the very different MC results for $S^3$, $\RP{3}$ and
$O(3)$,
it is even surprising that, for the $S^3$ model, the asymptotic prediction
does not fail for $\xi \simeq 10^8$ but works down to correlation lengths of
order $\xi \sim 10$.

Let us now speculate about the possible scenario in two dimensions.
{}From the above figures, it is very tempting to conclude to a phase transition
at a finite temperature
for all models with topological defects.
For models with boundary defects $\Pi_0 \neq 0$, similarly to the Ising model,
we have a linear $\beta$ function and a power law diverging correlation length.
A finite size scaling for the correlation length and specific heat
gives $\nu  = 0.78$  for $O(2)$ and $\nu = 0.71$ for $O(3)$.
Another possibility for this family of models would be to have two transitions,
the one at a higher temperature corresponding to the symmetry breaking
$O(n)\rightarrow SO(n)$.
As the broken symmetry is discrete, it can be accompanied with long range
order.
For models with point defects $\Pi_1 \neq 0$, similarly to the X-Y model,
the $\beta$ functions are compatible with a Kosterlitz-Thouless behavior $\beta
\sim |T-T_c|^{3/2}$.
Let us emphasize that, for model with topological defects, phase transitions at
a finite temperature are
completely compatible with a finite size scaling analysis of the data.
In contradistinction, the above RG predictions are consistent with the
simulation for models
with trivial topologies, but clearly ruled out for model with topological
defects.
Of course, the MC simulations and finite size scaling are always powerless
against arguments of the kind:
for a correlation length of order $\xi \sim 100$ (or $\xi \sim 10^{12}$)
there is a crossover to the predicted RG asymptotic behavior.
What is missing in order to draw a firmer conclusion is a theoretical basis.
We would like for example to have a modification of the Villain mapping for
models with curvature.
It is easy to build an energy-entropy argument for a pair of point defects,
but it seems difficult to turn this simple argument into a good approximation.
For a model believed to be in the same universality class as the $V_{3,2}$
model,
from an energy-entropy argument and from MC simulation,
Kawamura and Miyashita \cite{kawamura_84} already proposed a phase transition
at a finite temperature.
Beside possibly dangerous analogies with the X-Y model,
only the large $n$-limit for $\RP{n}$ concludes to a (first order)
transition at a finite temperature \cite{KZ92}.
Caracciolo {\em et all} \cite{caracciolo}, studying a mixed $S^3-\RP{3}$ model,
concluded also that the usual RG computation does not describe the $\RP{3}$
model
and proposed some interesting alternatives,
although still favoring a transition at $T_c = 0$ for the pure $\RP{3}$ model.
Another possible scenario would be that the naive power counting used to
compute the relevance of various term is valid only very close to $T=0$.
At higher temperature, the present results indicate that the least irrelevant
terms are the one related to topological defects.

Simply stated, in the range of temperatures accessible to MC simulations and by
using finite size scaling,
our data for all models with topological defects are very well described by a
phase transition
at a finite temperature.

I would like to thanks B.~Halperin, D.~Nelson and U.~T\"{a}uber for various
discussion.
This work is supported by the Fonds National Suisse de la Recherche
Scientifique, by the NSF grant DMR-91-15491 and by the office of Naval Research
grant N00014-93-1-0190.

\section*{Figures captions}
\begin{description}
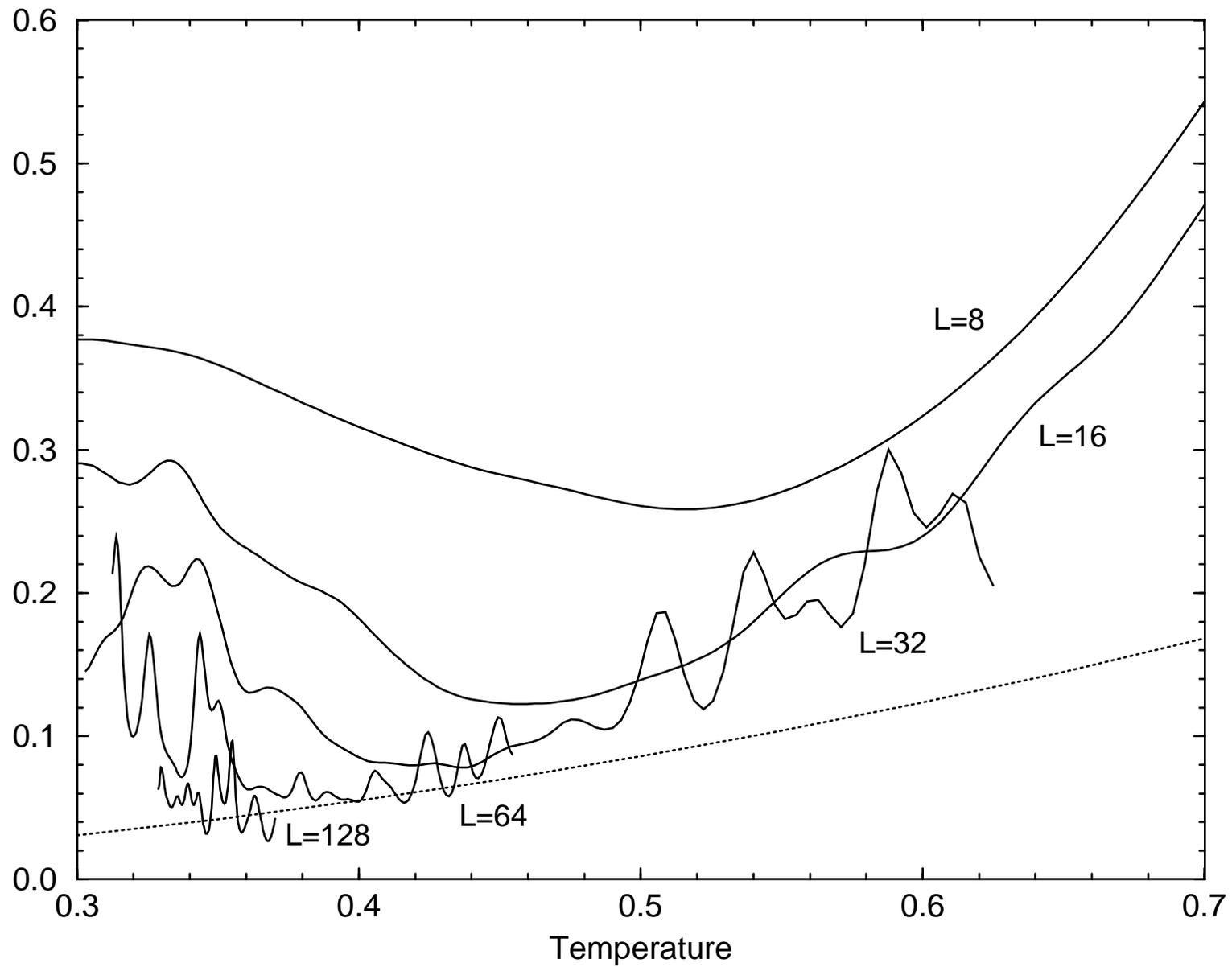
 \item[Figure 1:] The theoretical (dotted curve) and measured $\beta$ function
for the $V_{5,2}$ model.
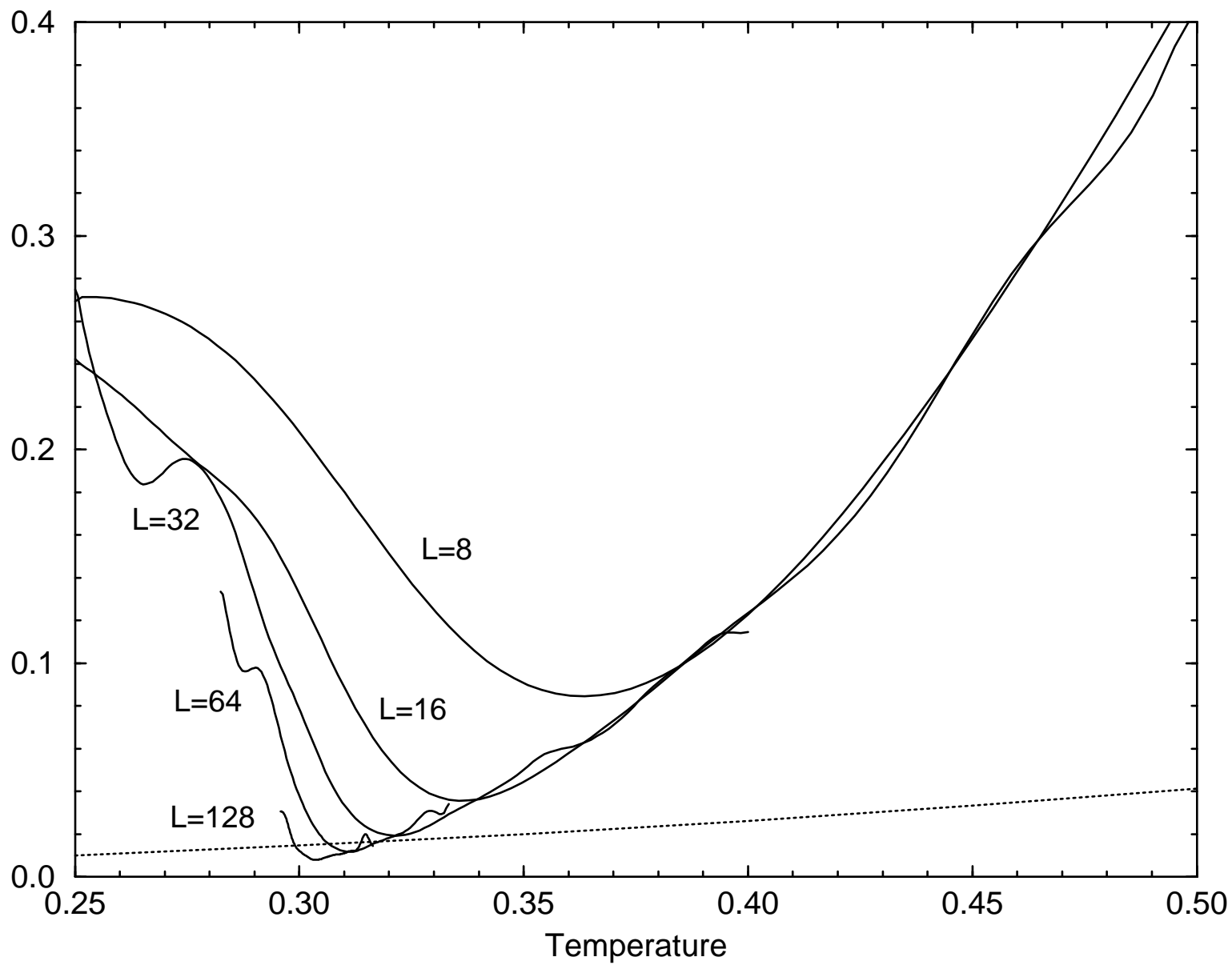
 \item[Figure 2:] As figure 1, but for the $\RP{3}$ model.
 At this scale, the one and five loops theoretical results are
undistinguishable.
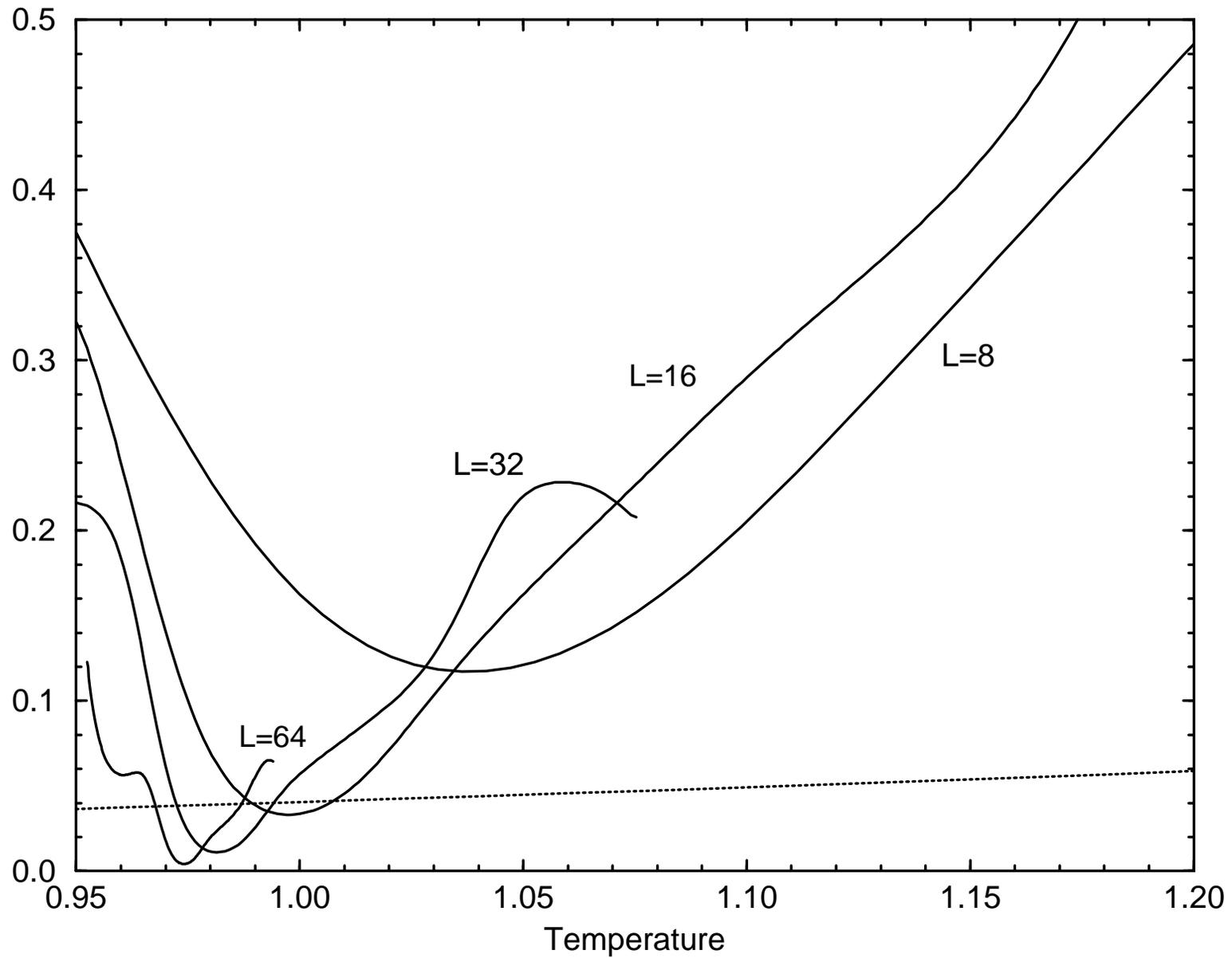
 \item[Figure 3:] As figure 2, but for the $O(3)$ model.
\end{description}

\end{document}